
\input harvmac
\input epsf
\ifx\epsfbox\UnDeFiNeD\message{(NO epsf.tex, FIGURES WILL BE IGNORED)}
\def\figin#1{\vskip2in}
\else\message{(FIGURES WILL BE INCLUDED)}\def\figin#1{#1}\fi
\def\ifig#1#2#3{\xdef#1{fig.~\the\figno}
\goodbreak\midinsert\figin{\centerline{#3}}%
\smallskip\centerline{\vbox{\baselineskip12pt
\advance\hsize by -1truein\noindent{\bf Fig.~\the\figno:} #2}}
\bigskip\endinsert\global\advance\figno by1}
\Title{\vbox{\baselineskip12pt
\hfill{\vbox{
\vbox{\hbox{\hfil EFI-95-03}\hbox{hep-th/9501039}}}}}}
{\vbox{\hbox{\centerline{On Consistent Boundary Conditions}}
\hbox{\centerline{for $c=1$ String Theory}}}}
\centerline{Martin O'Loughlin}
\smallskip
\centerline{Enrico Fermi Institute}
\centerline{University of Chicago}
\centerline{5640 S. Ellis Ave.}
\centerline{Chicago, IL 60637 USA}
\centerline{\tt mjol@yukawa.uchicago.edu}
\bigskip
\baselineskip 15pt plus 1pt minus 1pt
\bigskip
\noindent
We introduce a new parametrisation for the Fermi sea of the $c = 1$ matrix
model. This leads to a simple derivation of the scattering matrix,
and a calculation of boundary corrections in the
corresponding $1+1$--dimensional string theory. The new
parametrisation involves relativistic chiral fields, rather
than the non-relativistic fields of the usual formulations.
The calculation of the boundary corrections, following recent
work of Polchinski, allows us to place restrictions on the boundary
conditions in the matrix model. We provide a consistent set of
boundary conditions, but believe that they need to be
supplemented by some more subtle relationship between the
space-time and matrix model. Inspired by these
boundary conditions, some thoughts on the
black hole in $c=1$ string theory are presented.

\Date{January 1995}
\lref\mooreginsp{P. Ginsparg and G. Moore, ``Lectures on 2D Gravity
and 2D String Theory'' in {\it Recent Directions in Particle
Theory}, Proceedings of TASI 1992, editors, J. Harvey and
J. Polchinski, hep-th/9304011.}
\lref\polnewone{M. Natsuume and J. Polchinski, {\it Nucl.Phys.}{\bf B424},
(1994),137, hep-th/9402156.}
\lref\polnewtwo{J. Polchinski, ``On the Non-perturbative
Consistency of $d=2$ string theory'', hep-th/9409168,
NSF-ITP-94-96 (1994).}
\lref\wittzwie{E. Witten and B. Zwiebach, {\it Nucl. Phys.} {\bf B377},
(1992), 55, hep-th/9201056.}
\lref\kutadifr{P. Di Francesco and D. Kutasov, {\it Phys.Lett.}
{\bf 261B}, (1991), 385; {\it Nucl.Phys.} {\bf B375}, (1992), 119.}
\lref\polscatt{J. Polchinski, {\it Nucl.Phys.}{\bf B362}, (1991), 125.}
\lref\moorepless{G. Moore and R. Plesser, {\it Phys.Rev.}{\bf D46},
(1992), 1730, hep-th/9203060.}
\lref\polminzhng{D. Minic, J. Polchinski and Z. Yang,
{\it Nucl.Phys.}{\bf B369}, (1992), 324.}
\lref\mpr{G. Moore, R. Plesser and S. Ramgoolam, {\it Nucl.Phys.}
{\bf B377}, (1992), 143, hep-th/9111035.}
\lref\dmp{R. Dijkgraaf, G. Moore and R. Plesser, {\it Nucl.Phys.}
{\bf B394}, (1993), 356, hep-th/9208031.}
\lref\witt{E. Witten, {\it Phys.Rev.}{\bf D44}, (1991), 314.}
\lref\msw{G. Mandal, A. Sengupta and S. Wadia,
{\it Mod.Phys.Lett.}{\bf A6}, (1991), 1685.}
\lref\moore{G. Moore, {\it Nucl.Phys.}{\bf B368}, (1992), 557.}
\lref\cghs{C. Callan, S. Giddings, J. Harvey and A. Strominger,
{\it Phys.Rev.}{\bf D45}, (1992), R1005.}
\lref\polreview{J. Polchinski, {\it What is String Theory?}.}
\lref\mssf{G. Moore and N. Seiberg, {Int.J.Mod.Phys.}{\bf A7},
(1992), 2601.}
\lref\dvv{R. Dijkgraaf, E. Verlinde and H. Verlinde,
{\it Nucl.Phys.}{\bf B371}, (1992), 269.}
\lref\polold{J. Polchinski, {\it Nucl.Phys.}{\bf B362}, (1991), 125.}

\newsec{Introduction}

One of the current outstanding problems in theoretical physics is
the detailed understanding of quantum processes that involve black holes.
It is intriguing that we have a potential laboratory for studying these
processes, the $c=1$ matrix model, but frustrating that since
the discovery of this model no significant progress has been
made in studying its black hole physics.

We present here an alternative picture, to those generally
discussed \mooreginsp, of the relationship between the free fermions
of the matrix model and the space-time tachyon of two-dimensional
string theory.
As the fields we use to describe the matrix model are relativistic they
provide a more direct relationship between the matrix model and
the relativistic spacetime physics. In spirit, our
parametrisation is closest to that discussed in \wittzwie.
We rely for our intuition on the transform between the
matrix model and the string theory recently discussed by
Polchinski and Natsuume \polnewone\foot{This transformation is
motivated by comparing calculations in \refs{\moorepless,\polold}\
and  \kutadifr, but the explicit mapping
between field equations in the matrix model and in the
string theory was not discussed until the above mentioned
paper by Polchinski and Natsuume.}.

\newsec{Fermi Sea in Light Cone Co-ordinates}

The fermions of the $c=1$ matrix model arise from diagonalising
the matrix of the matrix model, taking a large N (dimension of the
matrix) limit and finding the critical scaling of coupling with
N such that in this limit the Feynman diagrams of the matrix model
can be thought of as smooth Riemann surfaces of differing genus.
This continuum limit can be described by the
scattering problem for non-relativistic
fermions in an inverted harmonic oscillator potential.

\eqn\norellag{
S = \half \int (\dot{\lambda^2} - \lambda^2) dx,\quad \dot{\lambda} =
{d\lambda \over dx}
}

The standard approach to this theory, at the classical level, is to
look at the low energy excitations of the fermi sea. These bosonic
excitations, the collective field, lead to a non-linear relationship
between left and right moving excitations - the classical S-matrix.
We will follow a similar approach, but the collective fields we will
use are different as we will now describe.

\ifig\fone{Co-ordinate systems
used throughout the paper and conventions for ingoing and
outgoing fermion wavefunctions. Curved arrows show
Hamiltonian flow on phase space.}{\epsfbox{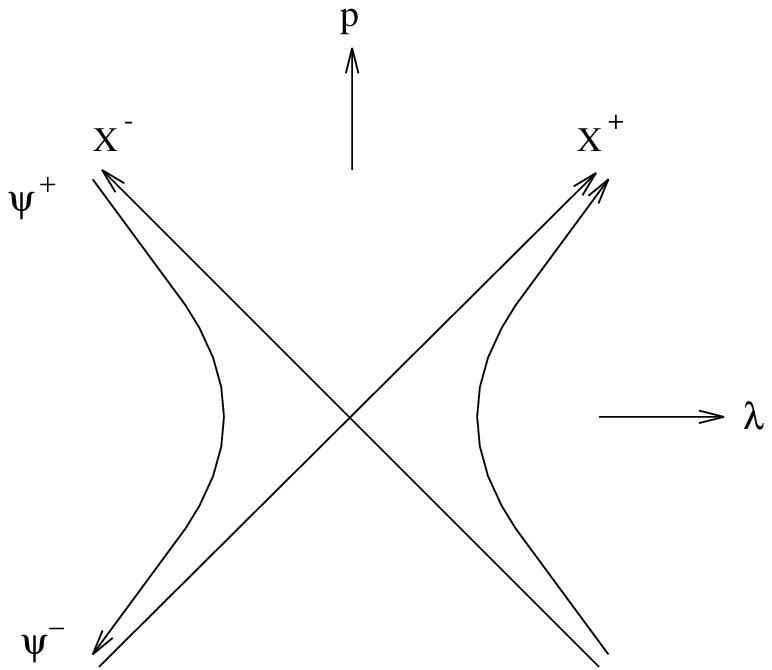}}

With $\dot{\lambda} = p$ let $X^\pm = p \pm \lambda$, (see \fone\
for our conventions). The classical
equation of motion is $\dot{p} = \lambda$ and the general solution
is $X^\pm = a_\pm(\sigma) e^{\pm t}$. For comparison with
the parametrisation of the Fermi sea in terms of $p_\pm(\sigma,t)$
\polold,
we will present a simple derivation of the classical S-matrix for
low energy scattering. Assuming the ingoing disturbance of the
Fermi sea never crosses the potential barrier, (which is sufficient
for this classical calculation), we can write,
$a_\pm(\sigma) = \pm a(\sigma) e^{\pm b(\sigma)} = \pm a(\sigma)
e^{\pm \sigma}$, and $h_\pm(\sigma) = log\ a(\sigma) \pm \sigma$.

For Fermi seas which obey the restriction that $X^+(X^-)$ is
monotonic, $h_\pm(\sigma)$ are invertible. Then

\eqn\chiral{\eqalign{
\chi_\pm (X^\mp, t) &\equiv (-p^2 + \lambda^2 - \mu) \cr
    &= (a^2(\sigma) - \mu) \cr
    &= a^2(h_\mp^{-1}(x^\mp \pm t)) - \mu
}}
where $x^\pm = log(\mp X^\pm)$. So we see that $\chi(v)$ are
chiral fields, and one can show that they trivially satisfy
$\chi_+(h_+(v)) = \chi_-(h_-(v))$. With a little more work
one can also show that $\chi_+(w) = \chi_-(-w + log(-\mu +
\chi_+(w)))$. This is the functional relationship between the
ingoing and outgoing collective field that leads to the
classical S--matrix of Polchinski and of Moore and Plesser\moorepless.
Putting,
\eqn\mmmodes{
\chi_\pm = \partial_\mp\bar{S}_\pm(u^\mp)
= \int_{-\infty}^\infty {d\omega \over
2\pi} {1\over \sqrt{2}} \bar{\alpha}_\pm(\omega)
e^{i\omega u^\mp},}
the recursive functional relationship between $\chi_\pm$
leads to a non-linear relationship between the modes $\bar{\alpha}_\pm$
which is the tree level collective field S--matrix.

If $\alpha_\pm$ are the creation and annihilation operators for the
$1+1$--tachyon, then comparing the calculations of Kutasov and DiFrancesco
to the tree level collective field S--matrix \refs{\moorepless,\polold},
the appropriate relationship between $\alpha_\pm$ and
$\bar{\alpha}_\pm$ is
\eqn\goverg{
\bar{\alpha}_\pm(\omega) = \alpha_\pm(\omega)
 ({\pi \over 2})^{\mp {i\omega\over 4}}
{\Gamma(\mp i\omega)\over \Gamma(\pm i\omega)},}
and combining this transformation with the collective field
S--matrix gives the classical tachyon S--matrix.

Another attractive feature of this
decomposition of the Fermi sea excitations is that there is
no breakdown of the collective field due to formation of folded
configurations as the sea evolves in time \polminzhng\ which in
the extreme case involves pulses propagating over the barrier.
To see the conservation of folds, define a fold
be a place at which the collective field can no longer be well
defined (it may become multiple valued). In our variables, this would
appear as a turning point in $X_\pm(\sigma,t)e^{\mp t}$, as a function of
$\sigma$. As these two functions are functions of
$\sigma$ only, the number of folds is encoded in $a(\sigma)$
alone and is clearly conserved.

\newsec{Canonical transformations of the Fermi Sea}

Even though this ``light cone'' description of the collective excitations
of the Fermi sea is very simple as described in the
previous section, to calculate higher order corrections to this
picture it is by far most convenient to go back
to the free fermion picture of the $c=1$ matrix model.  In this
picture the description of scattering is obtained by a
combination of fermionisation, free-fermi scattering and
bosonisation \mpr. The Feynman rules for the free fermi scattering
consist of a wall vertex given by the non-relativistic reflection
coefficient $R(x)$, and a free fermi propogator. Higher loop corrections
are given by sums of ring diagrams and are discussed in detail
in \mpr.

The same calculations may be carried
out here, but first we consider a simple method to derive the scattering
coefficient $R(x)$, the canonical transformation that
relates $\psi_+$ to $\psi_-$ (the ingoing and outgoing
fermion wavefunctions). Again this is a globally defined
transformation and thus we do not have to look at asymptotics to
work out the relationship between the wall scattering and the
relativistic fields at infinity (compare \refs{\moorepless,
\polold}).

As the coordinates $(X^+, X^-)$ are canonically conjugate,
the non-relativistic wavefunctions, $\psi_+$ and $\psi_-$,
are the Fourier transforms of each other, (this is exactly as
happens in non-relativistic quantum mechanics where the
Fourier transform relates the position and momentum representations
of a wavefunction).

\eqn\cantran{
\psi_-(X^+) = {1\over \sqrt{2\pi}} \int_{-\infty}^\infty\ dX^-
\ e^{iX^+X^-}\psi_+(X^-)}

The Schr\"odinger equation becomes
${1\over 4}(X^+X^- + X^-X^+) \psi_\pm = (i\partial_t - \mu)\psi_\pm$.
The normal ordering is the unique choice that ensures that
both left and right moving wavefunctions are delta function
normalisable.

For $\psi_+$ we find $({i\over 2} - i X^- \partial_- + \mu)
\psi_+^k(X^-) = k \psi_+^k(X^-)$.
The wavefunctions are,
\eqn\Lwvfns{
\psi_+^k(X^-) = a_k (X^-)^{-i(k - \mu) - {1\over 2}} \theta(X^-) +
b_k (-X^-)^{-i(k - \mu) - {1\over 2}} \theta(-X^-)}
and similarly for $\psi_-^k(X^+)$.

Applying the canonical transform to $\psi_+^k(X^-)$ we find
\eqn\Rwvfn{
\psi_-^k(X^+) = \cases{{\Gamma({1\over 2}-i(k - \mu))\over {\sqrt{2\pi}
(X^+)^{{1\over 2}-i(k - \mu)}}}
(a_k e^{i{\pi\over 2}({1\over 2}-i(k - \mu))}
+ b_k e^{-i{\pi\over 2}({1\over 2}-i(k - \mu))})& $X^+ > 0$\cr
{\Gamma({1\over 2}-i(k - \mu))\over {\sqrt{2\pi}
(-X^+)^{{1\over 2}-i(k - \mu)}}}
(a_k e^{-i{\pi\over 2}({1\over 2}-i(k - \mu))}
+ b_k e^{i{\pi\over 2}({1\over 2}-i(k - \mu))})& $X^+ < 0$}}
{}From these expressions we see that the reflection coefficient is given
by
\eqn\refcoeff{
R(k - \mu) = {\Gamma(\half -i(k - \mu))\over \sqrt{2\pi}} (
e^{-i{\pi\over 2}({1\over 2}-i(k-\mu))} + {b_k\over a_k}
e^{i{\pi\over 2}({1\over 2}-i(k-\mu))})
}
The boundary conditions are parametrised by the choice of
${b_k\over a_k}$, ($a_k \neq 0$ for the scattering problems considered
here). To relate these calculations to those with a boundary
at a fixed value of $\lambda$, (as in \refs{\moore, \mpr}), let us
consider in our framework the boundary conditions for which the left
and right moving fermion wavefunctions have the same form.
This amounts to requiring that
\eqn\eqform{
{a_k e^{-i{\pi\over 2}({1\over 2} - ix)} +
b_k e^{i{\pi\over 2}({1\over 2} - ix)} \over
{a_k e^{i{\pi\over 2}({1\over 2} - ix)} +
b_k e^{-i{\pi\over 2}({1\over 2} - ix)} }} = {b_k\over a_k}.
}
It is easy to show that this implies $a_k = \pm b_k$.

For the case with $a_k = b_k$ we find
\eqn\evenco{
R_{I+}(x) = \sqrt{2\over \pi} \Gamma(\half -ix)
cos({\pi\over 2}(\half -ix))}
and when $a_k = -b_k$
\eqn\oddco{
R_{I-}(x) = -i\sqrt{2\over \pi} \Gamma(\half -ix)
sin({\pi\over 2}(\half -ix)).}
The second (odd) case, is identical (to a phase) to the result
of \mpr\ for a wall at $\lambda = 0$.

For later purposes we will also write down the reflection
coefficient for the no wall scattering in terms of
$\psi_\pm$. This means there is no incoming wave from the
right hand side of the barrier, or in other words $\psi_+ = 0$ for
$X^- < 0$. Then
\eqn\nowallco{
R_{II}(x) = {1\over \sqrt{2\pi}} \Gamma(\half -ix)
e^{-i{\pi\over 2}({1\over 2}-ix)}
}

Notice that
the relativistic free fermions that we are describing,
can be exactly bosonised. In the language of
\dmp\  the S--matrix is related to a bosonisation of
fermionic Bogoliubov transformations on the in and
out states of the Fermi sea. In the discussion of
\mpr, the relativistic bosons that arise are found
in the asymptotic behaviour of the collective field. In our description
the bosons are everywhere relativistic.
In the fermion field theory the S-matrix may be modified by
choosing different in and out vacua around
which one considers scattering. It would be intriguing to
find some relationship between such choices of vacuum states for the
fermions, choices of vacua for the bosons, and thus possibly to
vacuum states for
quantum fields in flat or curved spacetimes.
{}From quantum field theory in curved space--times, we know that
understanding vacuum states of the field involves understanding the
relationship between wavefunctions in different asymptotic regions.
It is thus necessary for us to understand the relationship
between asymptotic regions in the matrix model and in the
space--time. We will say a little more about this at the
end of the next section, after we have discussed consistency of
boundary conditions in the matrix model.

\newsec{Low energy string theory}

We can now easily write down the expression for the full quantum
$1\to n$ amplitudes of the matrix model collective field
with general boundary conditions. One inserts the
appropriate reflection coefficient as computed in the
previous section, in
the formulae derived in \mpr.  For example, the
expression for the $1\to 2$ amplitude is

\eqn\onetwo{\eqalign{
\sqrt{3}\mu^{-i\omega}S(\omega;\omega_1, \omega_2) &= (\int_0^{\omega_1} -
\int_{\omega_2}^\omega) dx\ R(x - \mu) R(\mu + \omega - x) \cr
& \equiv f_\mu(\omega, \omega_1)
}}

We can use $f_\mu(\omega,\omega_1)$, adapting a calculation of
Polchinski and Natsuume \polnewone, to derive the second order
correction to the outgoing field in terms of the ingoing field. We will
sketch the outline, (the full details are well explained in \polnewone).

\eqn\second{\eqalign{
S_-^{(2)}(x^+) & = \int_{-\infty}^\infty {d\omega\over 2\pi}\
{1 \over \sqrt{2} i\omega} \int d\omega_1 f_\mu(\omega,\omega_1)
e^{i\omega x^+} {\Gamma(-i\omega)\over \Gamma(i\omega)}
 {\Gamma(-i\omega_1)\over \Gamma(i\omega_1)}
 {\Gamma(-i(\omega -\omega_1))\over \Gamma(i(\omega - \omega_1))}\cr
& \sqrt{2} i \omega_1 \int dx_1^-\ e^{-i\omega_1 x_1^-} S_+(x_1^-)
\sqrt{2} i (\omega - \omega_1) \int dx_2^-\ e^{-i
(\omega - \omega_1) x_2^-} S_+(x_2^-) \cr
& = {e^{x^+} \over \pi^2 \sqrt{2}} \int dx_1^-\ dx_2^-\
S_+(x_1^-) S_+(x_2^-) \int d\omega_1\ f_\mu(-i, \omega_1)
e^{-i \omega_1 (x_1^- - x_2^-) - x_2^-}
}}
Here $x^\pm = t\pm\phi$, where $(t,\phi)$ are the string target
space co-ordinates.
The second equality was obtained from the first by
considering the limit as $x^+\to-\infty$ which is dominated by the
first pole of the integrand in the lower half plane. The leading order of
perturbation theory in $1/\mu$ at $\omega = -i$, gives us
$f_\mu = {1\over 2\pi}$. The integral over $\omega_1$
leads to a delta function in $x_1^- - x_2^-$ giving the tree level
correction,
\eqn\pncorr{
S^{(2)tree}_-(x^+)  = {e^{x^+} \over 2 \pi^2 \sqrt{2}}\int dx^-
S_+(x^-)^2 e^{-x^-}.
}

One can show that these expressions are the
same as one derives from low energy string theory in the limit
that $x^+ \to -\infty$, (again we refer the reader to \polnewone\
for more details).
In string theory at $x^+ \to -\infty$ we are actually
calculating in a weak tachyon perturbation theory, and
not a large $\mu$ expansion as one does in the matrix model.
One can see this by noticing that the expansion parameter in
\pncorr is $e^{2\phi}\partial_x S$. For early enough times
this quantity is always small regardless of the size of $\mu$.

Therefore there is potentially some mixing of the
non--perturbative physics of the matrix model,
(where the perturbation theory is in terms of $1/\mu$),
with the perturbative physics of the low energy string theory.
Such would be a direct result of the non--local nature of
the transformation
between these two theories arising from the $\Gamma/\Gamma$
transformation on the $\alpha(\omega)$ \goverg, (a Hankel transform
in position space).  For example,
a Gaussian tachyon pulse maps to a collective field excitation
that has a decaying exponential
early time behaviour, $\bar{S}(x^-) \sim e^{x^-}$\polnewone. In the
collective field of the matrix model, this means that part of this
ingoing excitation of the fermi sea spends an exponentially long time
near the quadratic turning point, (even though the
bulk of the pulse is centered on $x^- \to x_0$). Thus
the non--perturbative tunneling
rate is enhanced. Such enhancement can cause corrections to
the low energy effective action in the string theory. We now
proceed to calculate these corrections.

\subsec{Non-perturbative corrections to low-energy string theory.}

We will evaluate
\eqn\Sminus{
S_-(x^+) = {e^{x^+}\over \pi\sqrt{2}} \int dx_1^-\ dx_2^-
\ S_+(x_1^-)S_+(x_2^-)\int d\omega_1 f_\mu(-i, \omega_1)
e^{-i\omega_1\Delta - x_2^-}}
where Im $\omega_1 = -\half$ and $\Delta = x_1^- - x_2^-$.

Notice that for $R_{II}$\nowallco and $R_{I+}$\evenco, the integrand
in $f_\mu(-i, x)$ has a double pole at $x = \mu - {i\over 2}$,
which is a point through which the $\omega_1$ integration
contour is required to pass. To investigate the effect of this we
deform slightly away from $\omega = -i$ to $\omega = -i + i\epsilon$.
The double pole is now a pair of poles at $x = \mu - {i\over 2}$ and
$x = \mu - {i\over 2} + i\epsilon$.
Possible subtleties can now be anticipated to arise from the choice of
integration for the two contours that appear in $f_\mu$. For the region
of the $\omega_1$ integral near the poles, (the part of the $\omega_1$
integral near $\omega_1 = 0$ is already accounted for and gave rise
to \pncorr)
\eqn\muminmuI{
\int d\omega_1 f_\mu(-i,\omega_1) e^{-i\omega_1\Delta - x_2^-} =
\int d\omega_1 e^{-i\omega_1 \Delta} (\int_0^{\omega_1} -
\int_{-i}^{-i-\omega_1}) {dx \over (x-\mu+{i\over 2})
(x-\mu+{i\over 2}-i\epsilon)}}
and concentrating on the first term
\eqn\muminmuII{\eqalign{
\int d\omega_1 e^{-i\omega_1 \Delta}
\int_0^{\omega_1} & {dx \over (x-\mu+{i\over 2})
(x-\mu+{i\over 2}-i\epsilon)}\cr
& = e^{-i\mu\Delta - {\Delta\over 2}} \int_{-\infty}^\infty
d\nu e^{-i\nu\Delta} \int_0^{\nu-{i\over 2}+\mu}
{dx \over (x-\mu+{i\over 2}) (x-\mu+{i\over 2}-i\epsilon)} \cr
& \sim i e^{-i\mu\Delta-{\Delta\over 2}}(\int_{-\infty}^\infty d\nu
e^{-i\nu\Delta} \int_\kappa^{\epsilon\over 2} {dy\over
(\nu + iy)(\nu+iy-i\epsilon)} \cr
& - {2\pi\over\epsilon}\int_0^\infty d\nu e^{-i\nu\Delta})
}}
The approximation in the second line of this expression is
obtained by restricting the $x$ integration to the
region near the poles. The additional term that appears in the
last line of \muminmuII\ arises from pinching the $dx$ contours
as $\epsilon \to 0$. When $\omega_1 > \mu -{i\over 2}$
and $\omega_1 < -\mu -{i\over 2}$ the contours are pinched
between the two poles.
\ifig\ftwo{Deformation of the $x$ integration contour
around the pole at $\mu - {i\over 2} + i\epsilon$ for
$\omega_1 > \mu -{i\over 2}$.}{\epsfbox{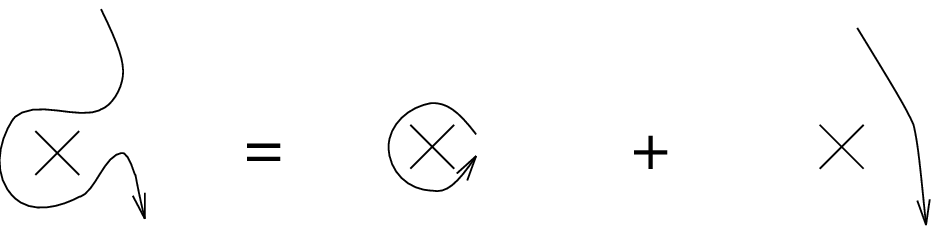}}
The deformation of the
contour required to navigate the poles for \muminmuII\
is shown in \ftwo.
By similar manipulations we find for the second term,
\eqn\muscnd{
i e^{i\mu\Delta-{\Delta\over 2}}(\int_{-\infty}^\infty d\nu
e^{i\nu\Delta} \int_{-\kappa}^{\epsilon\over 2} {dy\over
(\nu + iy)(\nu+iy-i\epsilon)}
- {2\pi\over\epsilon}\int_0^\infty d\nu e^{i\nu\Delta}).
}
We used the choice for all contours that one obtains
by taking the integrals along the real axis and continuously
deforming them into the lower half plane as $\omega\to -i$.

For this choice the final result has two terms on the
right hand side that are important for the early time
low energy string scattering. One term is that in \pncorr\
 and agrees with the low energy
string theory. The other, which comes from
the residues of the pole shown in the
first term on the right hand side of \ftwo,
and a similar pole from the other part of the integral,
is manifestly in disagreement with
the low-energy string theory. It is,
\eqn\deltas{
\delta S_-^{(2)}(x^+) = {2\pi\over \epsilon}\int dx_1^-\ dx_2^-
\ S^{(1)}_+(x_1^-)  S^{(1)}_+(x_2^-) {sin\mu\Delta\over\Delta}
e^{-\half(x_1^- + x_2^-)},}
where we still need to take the limit as $\epsilon\to 0$.  This
enables us to rule out this set of boundary conditions.

One may try alternative, though less natural, prescriptions for
the paths of integration, and though one may remove the
$1\over\epsilon$ divergence, one cannot remove an additional
finite non-local correction to the
correct second order low energy string theory.

Of our three reflection coefficients $R_{I+}, R_{I-}, R_{II}$,
only $R_{I-}$ is free from corrections in the second
order calculation, coming from double poles as discussed above.
However, for $\omega = -3i$, which appears
at higher orders in the weak field perturbation theory, one also
finds double poles in $R_{I-}(x)R_{I-}(\omega - x)$,
which the integration contour must negotiate.
These will contribute again to the perturbation theory in
a manner not in agreement with the known behaviour of the
low-energy string theory\foot{Although the tangle of contours
that one needs to sort out is greater for $\omega = -3 i$. We have
calculated corrections for $\omega = -2 i$, similar to those
discussed above, in
the perturbation expansion of $S_-^{(3)}$. Then only $R_{II}$ has double
poles that produce inconsistencies and  $R_{I\pm}$ have single poles at
$\omega = \mu - {i\over 2}$ that are potentially problematic.
For $\omega = -3 i$, $R_{I\pm}$ and
$R_{II}$ have several poles that will potentially produce additional
corrections to the low energy string theory.}, so $R_{I-}$
must also be ruled inconsistent.

We have not discussed other choices for $a_k/b_k$, or
other locations of the wall in $R(x, A)$. In both
cases the pole structure is more complicated than in the
simple symmetric cases we have discussed, and in general
are probably more problematic. For example, in the $A \to \infty$
limit of $R(x, A)$ the poles in the lower half plane accumulate
along the real axis. This will cause corrections to the
above calculations in the first step, where one takes
$x^+ \to -\infty$, as the first pole that one encounters in
the lower half plane will not be at $\omega = -i$ but at
Im $\omega > -1$.

As the problems arise from the poles in $R(x)$ for Im $x < 0$, we may
attempt to eliminate them by simply requiring boundary conditions for
which no such poles appear in the lower half plane.
Referring to \refcoeff, we see that this requires
${b_k\over a_k} = - e^{i\pi({1\over 2}-i(k-\mu))}$,which gives,
\eqn\nscoeff{
R_{NS}(x) = -i\sqrt{2\over\pi}\Gamma(\half - ix)\ sin\pi(\half -ix)
e^{i{\pi\over 2}({1\over 2}-ix)}.
}
We did not write down this reflection coefficient in our
discussion of boundary conditions in section 3, as we wanted
some symmetry between the ingoing and outgoing wavefunctions.
For this $R(x)$ we find that $\psi_-(x^+) = 0$ for $X^+ > 0$.
Using these boundary conditions is probably not the solution for
which we are searching, due to the lack of symmetry between ingoing
and outgoing wavefunctions. It
is amusing to note that this is closely related to the no wall
scattering coefficient \nowallco\
for which $\psi_+(x^-) = 0$ for $X^- < 0$.
Furthermore, $R_{NS}(x)$ appears as the
behaviour of $R_{I-}(x,A)$, as $A\to \infty$ at Im $x < 0$ \mpr\
(for Im $x > 0\ R_{I-} \to R_{II}$). This suggests that
maybe some less local identification between matrix
model phase space and space--time is required for
a consistent low--energy string theory. Similar
suggestions were made in \polnewtwo.

The most naive realization of this modified
mapping would be to use $R_{II}$ for ingoing fermions,
and $R_{NS}$ for the outgoing fermions, or vice versa, in the formulae
of \mpr. This will cause problems in the low energy string
theory as one may verify by calculating, similarly to the above,
the contribution from the logarithmic branch cut
in $f_\mu(-i, \omega_1)$, at $\omega_1 = \mu - {i\over 2}$,
(the source of the contribution is again the residue of the
pole that needs to be included due to the deformation of
the $x$ integration contour by the pole).
Another resolution would  simply be to use
$lim_{A \to \infty}R_{I-}(x,A) =
R_{II}^\prime(x).\ R_{II}^\prime(x)$ is singular only in the sense
that it is discontinuous across its branch cut along the
real axis. More significant than potential complications arising
from this branch cut, is the similarity between these
reflection coefficients and choices for bases of wavefunctions of
quantum fields in the presence of black holes. In such a case
one may consider {\it in} states that involve no component crossing the
past horizon (compare in $R_{II}$ with $\psi_+(X^-) = 0$ for $X^- < 0$),
and {\it out} states that involve no component crossing the future
horizon (compare in $R_{NS}$ with $\psi_-(X^+) = 0$ for $X^+ > 0$)
\foot{See for example \dvv, where a natural basis of vertex
operators consists of $\{ U_\omega^\lambda, V_\omega^\lambda\}$
where $U_\omega^\lambda$ vanish on the past horizon and $V_\omega^\lambda$
vanish on the future horizon.}. Taking this idea seriously suggests
that we should find Hawking radiation in the quantum tunneling
through the inverted harmonic oscillator, enhanced by the
Hankel transform that relates the matrix model and string theory
\ref\wip{M. O'Loughlin, {\it In Progress.}}.

\newsec{Conclusions}

The picture that we have developed here is unfortunately
still considerably removed from the picture of string
fields in space-time. To get the string amplitudes
correct to tree level, it is known that one needs to multiply the
collective field by a ratio of gamma functions, and that the space-time
tachyon then is a Hankel transform of the collective field.
This does not greatly enlighten us as
to the relationship between the matrix model fields and
the string fields. In particular, the main missing ingredient is
the bulk scattering of tachyons. The scattering that comes from
the reflection off the upside-down harmonic oscillator is
so-called wall scattering of tachyons.

However we do believe that using the fields that
we have introduced above, the relationship of the physics
of the Fermi sea to space-time
physics may be elucidated. The fields are manifestly relativistic
and thus are good candidates for fields in the matrix model
that have a simpler relationship to the vertex operators of the
space-time string theory.
It is intriguing that the decomposition implies a global
structure of free field {\it in} states, and free-field {\it out} states
with exact relativistic bosonisation.
These in and out fields are related simply by a canonical
transformation on the non-relativistic fermion wavefunctions
when expressed in the light-cone bases. The difficult fields
to find are the actual space-time string fields off-shell\foot{
Although with a slightly different definition of the macroscopic loop
operator one can find the ratio of gamma functions of \goverg\
directly in the fourier components of this redefined operator. In the
notation of Moore and Seiberg \mssf, we replace
$Tr(e^{-l M})$ with $Tr(J_1(e^{-l (\dot{M}\pm M)}))$.}.
The relationship to the macroscopic loop operators of
\mssf\ is not clear physically,  though one can use the
technique of canonical transformations to
write an expansion of the macroscopic loop operators in
powers of the in or out collective field modes. The formula
that one derives by this process is non--local and similar to the
momentum space form of the tree level string amplitudes.

Our fields also have a strong resemblance to the
in and out tachyon vertex operators that arise in the string theory
description of this model \kutadifr . In $1+1$-dimensional string theory
the tachyon vertex operator has the form,
\eqn\strtachyon{
T_k^\pm = e^{ikx \pm (ik-\sqrt{2})\phi},}
where $x$ is the flat direction and $\phi$ is the Liouville
direction.
This is suggestive of a relationship to the bosonisation of
our chiral fermion oscillators, for which the wavefunctions are
\eqn\fermwfn{
\psi \sim e^{ikt -ikx^- -{1\over 2}x^-}.
}

In this paper we have found that our light--cone decomposition
of the Fermi sea allows a simple discussion of boundary conditions,
and helps us to isolate a consistent set of such.
In discussing consistency conditions, Polchinski\polnewtwo\
reached similar
conclusions to ours on the subject of the boundary conditions
for $R_{I\pm},\ R_{II}$, and $R(x,A)$, in that they all
produce inconsistent space--time string physics. The details
of his arguments and ours do not agree. For example, we
do not find a key role played in our calculations by
the conserved charges $v_{mn}$, whereas these charges
form the basis for the conclusions in \polnewtwo. In the
light cone formulation where no walls appear, charges are
in general always conserved. Here the breakdown of the low energy string
theory arises through an interplay between the unstable
fermion--fermion resonances \mpr, where a fermion pair is
stuck at the top of the inverted oscillator potential,
and the Hankel transform.

The main challenges that lie ahead are to make explicit the exact
nature of the consistent non--perturbative completion of the matrix
model that we have presented, and to find more about relationships
between these boundary conditions and the
black hole. The resolution of the first appears to be related to
an identification of the relationship between the asymptotic
regions in the matrix model and string theory. Certainly an
understanding of
the black hole requires an understanding of asymptotic regions
(and horizons also). Our results, we believe, represent progress
in this direction.
\smallskip
\centerline{\bf Acknowledgements}
I would like to thank P. Freund, D.Kutasov and especially E. Martinec
for discussions. This work was supported by DOE grant DE-FG02-90ER40560.

\listrefs

\end